\documentclass[reqno,centertags
]{amsart}

\usepackage{amssymb}
\usepackage{enumerate}
\usepackage{amsmath}
\usepackage{amsfonts,mathrsfs}
\usepackage{amsthm}
 \usepackage[foot]{amsaddr}
\usepackage{latexsym,upref,enumerate}
\usepackage{nicefrac}
\usepackage{mathtools}
\usepackage{gensymb}
\usepackage{multirow}
\usepackage[pdftex,dvipsnames]{xcolor}  
\usepackage{color}
\usepackage{subcaption}
\usepackage{graphicx}
\usepackage{caption}
\usepackage{array}
\usepackage{hyperref}
\usepackage{nameref}
\usepackage[left= 1.2in, right= 1.2in]{geometry}
\usepackage{marginnote}
\usepackage{enumitem}
\usepackage[normalem]{ulem}
\usepackage[final]{microtype} 
\usepackage{stmaryrd}
\usepackage[percent]{overpic}
\usepackage[colorinlistoftodos,prependcaption,textsize=tiny]{todonotes}
\usepackage{xargs}

\newcommand*{\mailto}[1]{\href{mailto:#1}{\nolinkurl{#1}}}

\newcommand{\R}{\mathbb{R}}

\renewcommand{\O}{\mathcal{O}}

\newcommand{\e}{\varepsilon}
\renewcommand{\d}{\delta}
\renewcommand{\r}{\rho}
\renewcommand{\l}{\lambda}

\newcommand{\vp}{\varphi}
\renewcommand{\o}{\Omega}

\newcommand{\sech}{\mathrm{sech}}

\renewcommand{\t}{\bar t}
\renewcommand{\u}{\bar u}
\newcommand{\w}{\bar w}
\newcommand{\x}{\bar x}
\newcommand{\z}{\bar z}
\newcommand{\br}{\bar \rho}
\newcommand{\bo}{\bar \o}
\newcommand{\p}{\bar P}
\newcommand{\g}{\bar g}
\newcommand{\h}{\bar h}
\newcommand{\etab}{\bar \eta}

\renewcommand{\a}{\bar a}
\newcommand{\bl}{\bar \l}

\newtheorem{theorem}{Theorem}[section]

\theoremstyle{definition}

\newtheorem{remark}[theorem]{Remark}

\begin{document}

\title{Shallow water models for stratified equatorial flows}

\author{Anna Geyer$^1$}

\address{$^1$Delft University of Technology, Delft Institute of Applied Mathematics,
Faculty of EEMCS, Mekelweg~4, 2628 CD Delft, The Netherlands.}
\email{a.geyer@tudelft.nl}
\author{Ronald Quirchmayr$^2$}
\address{$^2$KTH Royal Institute of Technology, Department of Mathematics, Lindstedtsv\"agen 25
100 44 Stockholm, Sweden.}
\email{ronaldq@kth.se}

\maketitle
\begin{abstract}
Our aim is to study the effect of a continuous prescribed density variation on the propagation of ocean waves. More precisely, we derive KdV-type shallow water model equations for unidirectional flows along the Equator from the full governing equations by taking into account a prescribed but arbitrary depth-dependent density distribution. In contrast to the case of constant density, we obtain for each fixed water depth a different model equation for the horizontal component of the velocity field. We derive explicit formulas for traveling wave solutions of these model equations and perform a detailed analysis of the effect of a given density distribution on the depth-structure of the
corresponding traveling waves.
\end{abstract}

\section{Introduction}
\label{S-Intro}
\noindent
The dynamics of inviscid fluids are governed by Euler's equations of motion. In the context of water wave models one usually assumes the density function to be constant throughout the entire fluid body, which is a reasonable simplification since the compressibility of water is negligible: even very high amounts of hydrostatic or hydrodynamic pressure have almost no effect on the water density. 
In fact, the pressure acts like a Lagrange multiplier, enforcing
the incompressibility condition (see the discussion in~\cite{Con11}). However, other factors like temperature and salinity have a considerably higher effect on the density and give rise to stratified flows which become relevant e.g.~in the context of ocean dynamics. 
A significant increase of density due to changes in temperature or salinity gives rise to a pycnocline---a layer that separates surface water of lower density from water of higher density in greater depths. The propagation of disturbances of a pycnocline gives rise to internal waves, which are much larger than surface waves but  hard to detect since there is only weak interaction between internal waves and surface waves, cf.~\cite{FedorovBrown2009, CoJo15} and the references therein.  

In recent studies of stratified equatorial ocean flows, cf.~\cite{Constantin2012,  CoJo15,CoJo16}, the pycnocline has been considered as an infinitely thin interface between two layers of different but constant densities. 
This approach opened up rich developments in the theoretical investigations of equatorial flows, including
the derivation of exact solutions \cite{Constantin2012e, Constantin2013, Constantin2014a}, a generalization to three-dimensional fluid flows \cite{CoJo17},  discovery of the Hamiltonian structure of the model \cite{Constantin2015d, CIM16, Ionescu-Kruse2017a, Ivanov2017} and qualitative analysis of the thermocline see e.g.~\cite{Martin2015}; see also the overview articles~\cite{Johnson2015, Johnson2018} and  the reference therein.  A comprehensive account of phenomena pertaining to physical oceanography in general and to equatorial ocean dynamics in particular are presented in \cite{CoJo16b}. 

While the above approach assumes a predefined \emph{discontinuous} piecewise constant density function with a jump at the pycnocline, the aim of the present paper is to study the effect of a \emph{continuous change of density} on the propagation of water waves.
To this end we derive  the weakly nonlinear KdV-type shallow water equation 
\begin{align*} 
2 u_{\tau} - 2\o_0 u_{\xi} + \nu u u_{\xi} + \frac{1}{3} u_{\xi\xi\xi} =0
\end{align*}
for prescribed but arbitrary depth-dependent density distributions.
Here the unknown $u$, depending on time $\tau$ and space $\xi$, describes the horizontal velocity component of the flow field beneath a unidirectional free surface shallow water wave of small amplitude at a specific water depth; the coefficient $\nu$ of the nonlinear term $u u_\xi$, to be specified in the sequel,  depends on the prescribed density distribution and the water depth.
The major novelty compared to the case of constant density is the depth dependence of the horizontal velocity component of the flow at leading order. In particular, the vorticity of these flows does \emph{not} vanish, cf.~Remark \ref{rem-vorticity}. This is expected in view of the general principle that stratified flows can never be irrotational, see the discussion in \cite{Con11} and \cite{Walsh09}. 
The constant $\o_0$ incorporates the Coriolis effect along the Equator. We obtain a similar equation for the elevation of the free surface.
By a thorough study of the explicit traveling wave solutions of these KdV type equations we acquire a solid theoretical understanding of the effect of the density distribution on the size and shape of these traveling waves at various depths. This analysis shall serve as a basis for a better understanding on how a continuous change of density affects the dynamics of more general types of waves.


\section{Governing equations in dimensionless form}
\label{S-fplane}
\noindent
In this section we introduce the governing equations for unidirectional gravity water waves; we provide the equations in physical variables and bring them into dimensionless form in the sequel. The main difference compared to the classical water waves problem is that we allow for a non-constant  ``background'' density distribution depending smoothly on the vertical direction caused by e.g. increasing salinity or decreasing temperature with depth. Such stratification phenomena occur in particular in equatorial regions, where waves typically propagate in purely azimuthal direction, so that a unidirectional setting is a reasonable simplification, cf.~\cite{LeBlond1978} and the discussion in  \cite{CoJo15}. 
Motivated by this geophysical point of view, in which the effects of the earth's rotation on large waves become relevant, we consider the $f$-plane approximation for the inviscid Euler equations for two-dimensional unidirectional flows near the Equator.
The motion of the water between the flat bed and the free surface is therefore governed by the following set of equations:
\begin{align} 
\begin{aligned}
\label{eq-gov}
\u_{\t} + \u\u_{\x} + \w \u_{\z} + 2\bo \w &= -\br^{-1} \p_{\x} \\
\w_{\t} + \u\w_{\x} + \w\w_{\z} - 2\bo  \u &= -\br^{-1} \p_{\z} - \g \\
\br_{\t} + (\br \u)_{\x} + (\br  \w)_{\z} &= 0, 
\end{aligned}
\end{align}
where the bars distinguish these physical variables from dimensionless variables appearing later. The independent variables $\x$ and $\z$ denote the directions of increasing azimuth and vertical elevation respectively; the horizontal and vertical fluid velocity component in the direction of increasing azimuth and elevation are given by  $\u =\u(\x,\z,\t)$ and $\w=\w(\x,\z,\t)$. The pressure is denoted by $\p=\p(\x,\z,\t)$. 
We consider a prescribed but arbitrary continuously differentiable density distribution $\br$ of the water which may vary with depth only, that is 
\begin{equation}
 \br  = \br(\z)>0.
\end{equation}
\begin{figure}[h!] \centering
\begin{subfigure}{.53\textwidth}  \centering
\begin{overpic}[width=.95\textwidth]{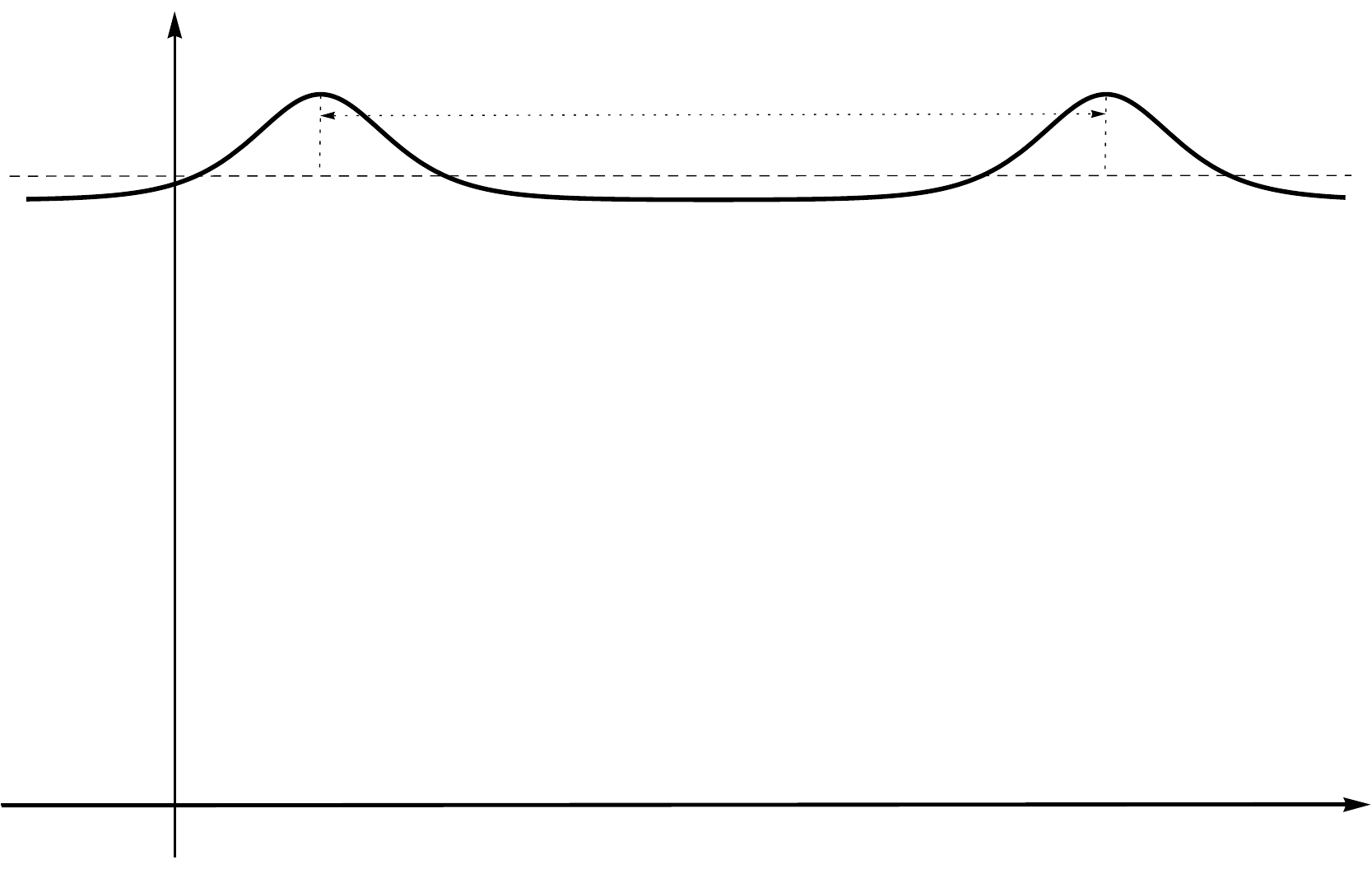}
      \put(0,5.5){\small{$\z=0$}} 
      \put(97,0.5){\small{$\x$}}
      \put(0,52){\small{$\z=\h_0$}}  
      \put(20.5,52){\small{$\a$}} 
      \put(50,56){\small{$\bl$}} 
      \put(79,58){\small{$\h_0 + \etab(\cdot,\t)$}}      
\end{overpic}
\caption{}
\label{fig:fluid_dom}
\end{subfigure} \quad
\begin{subfigure}{.21\textwidth}  \centering
\begin{overpic}[width=.96\textwidth]{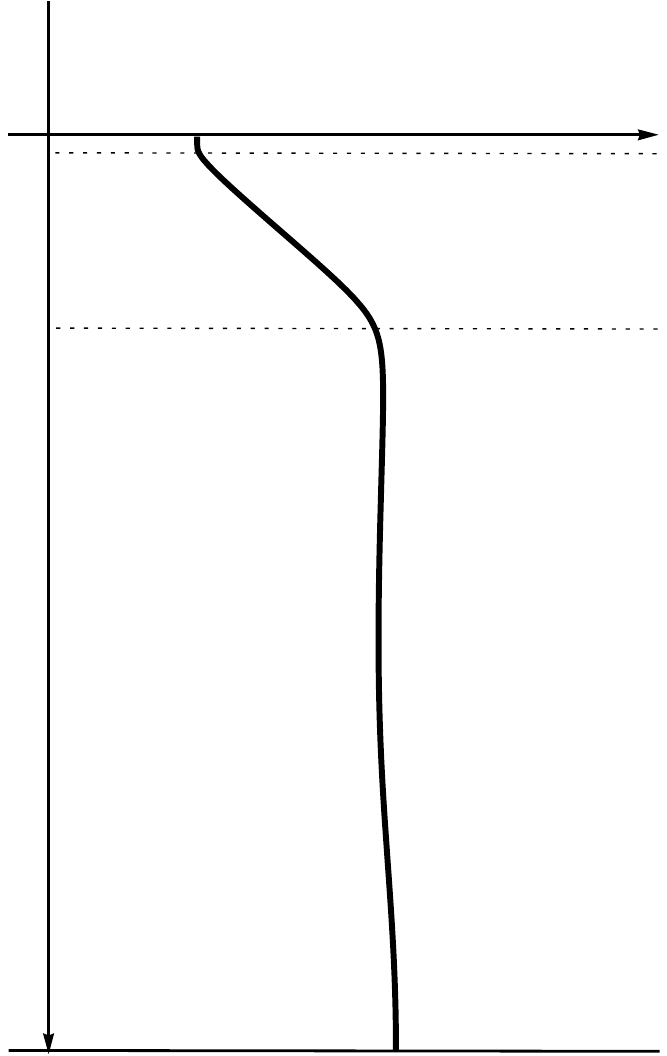}
	\put(7,2.5){\small{$\z=0$}}
	\put(7,89){\small{$\z=\h_0$}} 
	\put(37,40){\small{$\br(\z)$}}
	\put(34,76.5){\tiny{pycnocline}}
\end{overpic}
\caption{}
\label{fig:density}
\end{subfigure}
\caption{Fig.~\ref{fig:fluid_dom} illustrates the fluid domain in the physical $(\x,\z)$-plane between the flat bed at $\z=0$ and the free surface $\z=\h_0+\etab(\cdot,\t)$ at a certain instant of time $\t$. The average water level $\h_0$ is indicated by a dashed line, $\bl$ shows the distance between two consecutive crests and $\a$ is the vertical deviation of a typical crest from $\h_0$. 
Fig.~\ref{fig:density} shows a prescribed depth dependent density distribution $\br(\z)$ with a significant density increase in the region between the two dotted horizontal lines close to the surface giving rise to a pycnocline.
} 
\label{fig:1}
\end{figure}
The gravitational acceleration is denoted by $\g\approx 9.81m\,s^{-2}$ and 
the constant $\bo~\approx~7.29~\cdot~10^{-5}~\mathrm{rad}\,s^{-1}$ denotes the rotational speed of the Earth around the polar axis. Thus, the two $\bo$-terms in \eqref{eq-gov} capture the effects of the so-called \emph{Coriolis force}. 
Let us emphasize at this point that all considerations we make in this paper equally apply to the corresponding non-geophysical settings, i.e.~to unidirectional inviscid gravity water waves with a prescribed smooth depth-dependent density distribution. In this case we simply set $\bo=0$.

As already mentioned, we assume that the fluid body is bounded from below by a flat bed at $\z=0$; it is bounded from above by the free water surface which is assumed to be described by the graph of the function  $\h_0 + \etab(\cdot,\t)$ at any instant of time $\t$, where $\etab(\x,\t)$ measures the deviation of the free surface from the average water depth $\h_0$ at $(\x,\t)$.
We impose the usual dynamic boundary condition for the pressure at the water surface and the kinematic boundary condition for the vertical velocity at the surface and the flat bed:
\begin{align} 
\begin{aligned} \label{eq-bd_cond}
\p=\p_{\mathrm{atm}} \quad &\text{on} \quad \z=\h_0 + \etab(\x,\t)  \\
\w=\etab_{\t} + \u\etab_{\x} \quad &\text{on} \quad \z=\h_0 + \etab(\x,\t)\\
\w=0 \quad &\text{on} \quad \z=0;
\end{aligned}
\end{align}
the constant $\p_{\mathrm{atm}}$ denotes the atmospheric pressure.  
 \smallskip
 
To bring  the set of equations \eqref{eq-gov} and boundary conditions \eqref{eq-bd_cond} into dimensionless form, we use the following standard reference length scales: a typical amplitude of the surface wave $\a$ and the average undisturbed water depth $\h_0$ as the vertical scales, and a typical wave length $\bl$ as the horizontal scale, cf.~\cite{Constantin2008k},~\cite{Johnson1997} and~\cite{Joh02} for more details. We introduce the following set of non-dimensional variables:
\begin{align}
\begin{split}
\label{eq-nondim}
    &\x= \bl x, \quad \z = \h_0 z, \quad  \t=\frac{\bl}{\sqrt{\g \h_0}} t,\quad
    \u= \sqrt{\g\h_0} u, \quad \w = \sqrt{\g\h_0} \frac{\h_0}{\bl} w, \quad \etab = \a \eta, \\
    &\p= \p_{\mathrm{a t m}} -  \g\int_{\h_0}^{\z}\br(s) \, \mathrm d s +\g\h_0\br P,  \quad       
    \bo = \frac{\sqrt{\g \h_0}}{\h_0} \o, 
\end{split}
\end{align}
where the dimensionless pressure $P$ acts as a measure for the deviation of $\p$ from the hydrostatic pressure distribution.
The density $\br$ is scaled according to 
\begin{equation}
\label{eq-densitynondim}
    \br (\z) = \br_0 \r(z), \text{ where } \br_0 = \br(\h_0).
\end{equation}
By employing \eqref{eq-nondim}--\eqref{eq-densitynondim} and introducing the two fundamental dimensionless parameters  
\begin{equation}
\label{eq-epsdel}
    \e \coloneqq \frac{\a}{\h_0},   \qquad \d \coloneqq \frac{\h_0}{\bl},
\end{equation}
referred to as \emph{amplitude} and \emph{shallowness} parameter, the set of equations \eqref{eq-gov} and boundary conditions \eqref{eq-bd_cond} transform into the following dimensionless form: 
\begin{align}
\begin{aligned}  \label{sys-eqnondim}
    u_t + uu_x + wu_z + 2\o w &= -P_x \\
    \d^2(    w_t + uw_x + ww_z) - 2\o u &= -\frac{(\r P)_z}{\r} \\
    u_x  + w_z +\frac{\r_z}{\r} w &=0\\
   \\
    P=\frac{1}{\r(z)}\int_1^z\r(s) \,\mathrm d s \quad &\text{on} \quad z=1 + \e\eta(x,t)  \\
    w=\e(\eta_t + u\eta_x) \quad &\text{on}  \quad z=1 + \e\eta(x,t) \\
    w=0  \quad &\text{on} \quad z=0.
    \end{aligned}
\end{align}
 In order to transfer the free surface to a fixed boundary, we rewrite the boundary conditions at the free surface by means of Taylor expansions of the involved variables $u$, $w$ and $P$ about $z=1$. This yields the expressions
\begin{align}
\begin{aligned} 
    P + \e \eta P_z +\e^2 \eta^2/2 P_{z z} 
    = \e \eta - \e^2\eta^2 \frac{\r_z}{2}  +\O(\e^3) \quad &\text{on} \quad z=1 \\
    w=\e(\eta_t + u\eta_x-\eta w_z) +\O(\e^2) \quad &\text{on}  \quad z=1 \\
    \end{aligned}
\end{align}
where we have used the fact that $\rho(1)=1$ in view of the nondimensionalization \eqref{eq-densitynondim}. 
Finally, following \cite{Constantin2008k}, we perform the usual scaling 
\begin{equation} \label{scale_u_w_P}
u\mapsto \e u, \; w \mapsto \e w, \; P\mapsto \e P
\end{equation}
on the system \eqref{sys-eqnondim} and set
\begin{equation}
\label{eq-scaledomega}
    \o = \e \o_0, 
\end{equation}
where $\o_0$ is an appropriate constant.
The latter scaling is a valid manoeuvre since the dimensionless parameters $\o$ and $\e$ are of the same order of magnitude when assuming typical values of ocean depth and wave amplitude for offshore waves near the Equator, see the discussion in~\cite{GeyQui2018}.
Additionally we exploit that the shallowness parameter $\d$ in system \eqref{sys-eqnondim} can be scaled out in  favor of $\e$ via the transformation
\begin{align}
\begin{split}
\label{eq-johnscale}
    &x\mapsto\frac{\d}{\sqrt{\e}}x,\quad z\mapsto z, \quad t\mapsto \frac{\d}{\sqrt{\e}}t, \\
    &P\mapsto P, \quad \eta \mapsto \eta, \quad u\mapsto u,\quad  w \mapsto \frac{\sqrt{\e}}{\d} w,
\end{split}
\end{align}
where the scaling of $w$ is required to ensure conservation of mass in the resulting system. We note that this transformation remains  non-singular for $\e$ and $\d$ tending to zero only if these parameters satisfy a certain asymptotic relation. That is, by imposing the transformation \eqref{eq-johnscale} one considers the restriction to \emph{shallow water waves of small amplitude}:
\begin{equation}
\label{eq-smallampregime}
     \qquad \d \ll 1, \qquad  \e= \O(\d^2). 
\end{equation}
The application of  \eqref{scale_u_w_P}, \eqref{eq-scaledomega} and \eqref{eq-johnscale} on the system \eqref{sys-eqnondim} yields the following system of governing equations for unidirectional equatorial waves in scaled and dimensionless form:
\begin{align}
\begin{aligned}  \label{sys-eqscalednondim}
    u_t + \e(uu_x + wu_z) + 2\e \o_0 w &= -P_x   \\
    \e(    w_t + \e(uw_x + ww_z)) - 2\e \o_0 u &= -\frac{(\r P)_z}{\r}   \\
    u_x  + w_z +\frac{\r_z}{\r} w &=0\\
    \\
   P= \eta - \e \eta \Big(  \frac{\eta}{2} \r_z - P_z\Big)  \quad &\text{on} \quad z=1  \\
        w =\eta_t + \e (u\eta_x-\eta w_z) \quad &\text{on} \quad z=1 \\
    w=0 \quad &\text{on} \quad z=0.
    \end{aligned}
 \end{align}

\section{Derivation of a shallow water model}
\label{S-KdV}
\noindent
The system \eqref{sys-eqscalednondim} of governing equations for equatorial waves serves as a starting point for our derivations. We follow the techniques presented in \cite{Johnson1997,Joh02} to obtain model equations for the surface elevation $\eta$ and the horizontal velocity $u$.
By retaining only the leading order terms of \eqref{sys-eqscalednondim} (i.e.~by setting $\e=0$) we obtain the linear system
\begin{align}
\begin{aligned}  \label{sys_leading_order_xt}
    u_t &= -P_x    \\
     0&  =(\r P)_z   \\  
    u_x & + w_z +  \frac{\r_z}{\r} w &=0\\    
    \\
    P = \eta \quad  &\text{on} \quad z=1     \\
    w =\eta_t  \quad &\text{on} \quad z=1 \\
    w=0 \quad &\text{on} \quad z=0,
\end{aligned}
\end{align}
from which we infer that the leading order approximation of $\eta$ (again denoted by $\eta$) satisfies the linear wave equation: 
\begin{equation}
\label{eq-linwaveeq}
    \eta_{t t} - \eta_{x x} = 0. 
\end{equation}
To see this, we first note that the second equation in \eqref{sys_leading_order_xt} and the dynamic boundary condition imply that $\eta=\r P$ for all $z\in[0,1]$. In view of the first equation this yields that $ \r u_t = - \eta_x$.
Next we invoke the equation of mass conservation which is equivalent to $(\rho w)_z=-\rho u_x$, hence
$$w(x,z,t)=-\r(z)^{-1}\int^z_0 \r(s) u_x(x,s,t) \, \mathrm d s.$$
 Differentiating both sides with respect to $t$,  using the  boundary condition at the top and the fact that $\r(1)=1$,  we obtain that  $\eta_{tt} =- \int_0^1 \r u_{xt}\, \rm d s$ on $z=1$. Hence, we infer that $\eta$ satisfies \eqref{eq-linwaveeq}.\medskip

The general solution of \eqref{eq-linwaveeq} is of the form $\eta(x,t)=F_{\shortrightarrow}(x-t)+F_{\shortleftarrow}(x+t)$ for arbitrary functions $F_{\shortrightarrow}$ and $F_{\shortleftarrow}$. 
This motivates the introduction of the \emph{far-field variables}
\begin{equation}
\label{eq-farfieldvars}
    \xi =x- t, \quad \tau = \e t
\end{equation}
to follow right-propagating waves governed by \eqref{sys-eqscalednondim}. Rewriting \eqref{sys-eqscalednondim} in terms of the far-field variables \eqref{eq-farfieldvars} yields the system
\begin{align}
\begin{aligned} \label{eq-eqsfarfield}
    -u_{\xi} + \e(u_{\tau} + uu_{\xi} + wu_z + 2 \o_0 w) &= -P_{\xi} \\
       \e (-w_{\xi} + \e( w_{\tau} +uw_{\xi} + ww_z) - 2 \o_0 u) &=-\frac{(\r P)_z}{\r}\\
    (\rho w)_z &= -\rho u_{\xi} \\
                \\
   P= \eta - \e \eta \Big(  \frac{\eta}{2} \r_z - P_z\Big)  \quad &\text{on} \quad z=1  \\
 w=-\eta_{\xi} + \e(\eta_{\tau}+ u\eta_{\xi} - \eta w_z) \quad &\text{on}  \quad z=1    \\
    w=0 \quad &\text{on} \quad z=0.
    \end{aligned}
\end{align}
To obtain an asymptotic solution of system \eqref{eq-eqsfarfield}, we formally expand the respective variables $\eta$, $u$, $w$ and $p$ in the form $q \sim \sum_{n=0}^{\infty}q_n \e^n$. 
At leading order we obtain the linear system
\begin{align}
\begin{aligned} \label{eq-eqsfarfield_e^0}
  u_{0 \xi}  &= P_{0 \xi}  \\
     0&  =(\r P_0)_z \\       
 (\rho w_0)_{z} & =-\rho u_{0\xi} \\
        \\
        P_0 = \eta_0 \quad &\text{on} \quad z=1    \\
            w_0 =-\eta_{0 \xi}  \quad &\text{on} \quad z=1 \\
w_0=0 \quad &\text{on} \quad z=0.
\end{aligned}
\end{align}
Integrating the second equation and using the boundary condition on the pressure we obtain that $\eta_0= \r P_0 $ for all $z\in [0,1]$. In view of the first equation we may therefore deduce that 
\begin{equation}
\label{eq-u-eta}
    u_{0} = \r^{-1}\eta_{0},
\end{equation}
where we have invoked the assumption that any change in $u$ can occur only due to a perturbation of $\eta$, see \cite{Johnson1997}. From the equation of mass conservation, the bottom boundary condition and \eqref{eq-u-eta} we deduce that $w_0 = - z \r^{-1}  \eta_{0\xi}$. 
 Next we consider the first order system
\begin{align}
\begin{aligned} \label{eq-eqsfarfield_e^1}
 - u_{1 \xi} + \r^{-1} \eta_{0 \tau} + \r^{-2}  (1+z \r^{-1} \r_z)\eta_0 \eta_{0 \xi} 
    - 2\Omega_0 z \r^{-1}\eta_{0 \xi}   &= -P_{1 \xi}  \\
             z \eta_{0 \xi \xi} - 2\Omega_0  \eta_0 &=-(\r P_1)_z  \\
 (\rho w_1)_{z}&=-\rho u_{1\xi} \\
            \\
      P_1= \eta_1 +  \frac{\r_z}{2} \eta_0^2 \quad &\text{on} \quad z=1    \\
 w_1=-\eta_{1\xi} + \eta_{0\tau}+ (2-\r_z)\eta_0\eta_{0\xi} \quad &\text{on}  \quad z=1    \\
    w_1=0 \quad &\text{on} \quad z=0,
\end{aligned}
\end{align}
where we have written all leading order expressions in terms of $\eta_0$.
Viewing the second equation as a linear ODE in $P_1$ with $z$ being the independent variable, we can solve it using the pressure boundary condition and obtain that  
\begin{equation*}
P_1(z) = \r^{-1}\left [P_1(1) +\int_1^z 2\Omega_0\eta_0 - s\eta_{0\xi\xi} \, \mathrm d s\right].
\end{equation*}
This results in the following representation of $P_1$ in terms of $\eta_0$ and $\eta_1$:
\begin{equation*}
P_1(z) =  \r^{-1}\left [\eta_1 + \frac{\r_z}{2} \eta_0^2 + 2\Omega_0 \eta_0(z-1) - \frac{z^2-1}{2}\eta_{0\xi\xi} \right ]
\end{equation*}
Next we solve the equation of mass conservation as a linear ODE in $w_1$. By expressing the inhomogeneity  using the first equation, and incorporating the boundary condition on $z=0$ and $P_{1\xi}$ obtained in the previous step, we find that 
\begin{align} \label{w_1_int}
w_1(z) &= -\frac{1}{\r}\int^z_0 \r u_{1\xi} \, \mathrm d s\notag\\
&=   -\frac{1}{\r}\int^z_0 \Big[  \eta_{1\xi} + \eta_{0\tau}  - 2\Omega_0\eta_{0\xi} - \frac{s^2-1}{2}\eta_{0\xi\xi\xi} + 
\Big(\r_z+\frac{1}{\r} + \frac{s \r_z}{\r^2} \Big)\eta_0\eta_{0\xi} \Big]\, \mathrm d s.
\end{align}
Finally, we  calculate the integral in \eqref{w_1_int}, evaluate the resulting expression for $w_1$ at $z=1$ and compare the outcome with the corresponding boundary condition in \eqref{eq-eqsfarfield_e^1}. This yields the weakly nonlinear third order partial differential equation 
\begin{equation} \label{eq-sgKdV}
    2 \eta_{\tau} - 2\o_0 \eta_{\xi} + \nu(\r) \eta \eta_{\xi} + \frac{1}{3} \eta_{\xi\xi\xi} =0,        
\end{equation}
where
\begin{equation}
\label{eq-nu}
    \nu(\r) =   2 -\r_z(1) +\int_0^1 \Big(\r_z+ \frac{(z \r)_z}{\r^2} \Big) \, \mathrm d z  \;\in \R,
\end{equation}
which governs the leading order approximation of the \emph{free surface $\eta$}, where $\eta \sim \eta_0+\O(\e)$.
The equation for the leading-order approximation of the \emph{horizontal  velocity  $u$} at any depth $z\in[0,1]$ has a depth-dependent coefficient of the nonlinear term: 
\begin{equation} 
\label{eq-sgKdVu}
2 u_{\tau} - 2\o_0 u_{\xi} + \r(z)\nu(\r) u u_{\xi} + \frac{1}{3} u_{\xi\xi\xi} =0.
\end{equation}

\begin{remark}
\label{rem-transftostdKdV}
For constant density $\rho \equiv 1$  and vanishing Coriolis force, both equations \eqref{eq-sgKdV} and \eqref{eq-sgKdVu}  reduce to the  standard KdV \cite{Korteweg1895}
\begin{equation}
\label{eq-stdKdV}
   2u_{\tau} + 3 uu_{\xi} + \frac{1}{3}u_{\xi\xi\xi}=0,
\end{equation}
which approximates right-propagating shallow water waves of small amplitude at leading order.  
We observe that for fixed $z\in[0,1]$ and for any given density function $\r$  the coefficient of the nonlinear term in both equations is a fixed real number. Therefore, these equations can always be brought to the form \eqref{eq-stdKdV} via the transformation 
\begin{equation}
\label{eq-transf-stdKdV}
U = (\alpha/3)^{3/5} u, \quad T =(\alpha/3)^{3/5}\tau, \quad 
X=(\alpha/3)^{1/5}(\xi + \Omega_0 \tau), 
\end{equation}
where $\alpha=\nu(\rho)$ and $u=\eta$ for \eqref{eq-sgKdV} and $\alpha=\rho(z)\nu(\rho)$ for \eqref{eq-sgKdVu}.  
This transformation is valid only as long as $\alpha>0$. The sign of $\alpha$ can be determined for any arbitrary given density distribution $\rho$; we note that $\alpha$ is strictly positive for realistic density distributions in the context of (geophysical) water waves and  provide examples for different densities at the end of Section \ref{s-tws}. 
\end{remark}
\begin{remark}
\label{rem-integrability}
In view of Remark \ref{rem-transftostdKdV} several remarkable properties of the standard KdV \eqref{eq-stdKdV}, like soliton interaction, bi-Hamiltonian structure, the existence of a Lax pair, integrability, global Birkhoff coordinates, see e.g.~\cite{Gardneretal67, Lax68, FaddZakh71, KappelerPoeschel03}, and stability of traveling wave solutions (cf.~Section \ref{s-stability} for some references regarding stability) remain valid for equations  \eqref{eq-sgKdV} and \eqref{eq-sgKdVu} by means of the transformation \eqref{eq-transf-stdKdV}. 
\end{remark}

\begin{remark}
\label{rem-vorticity}
The depth dependent coefficient in the  velocity equation \eqref{eq-sgKdVu} due to non-constant density $\rho$ represents a major difference compared to the standard KdV, which models both the free surface $\eta$ and the horizontal velocity $u$ at \emph{any} depth. In the next section we will consider explicit traveling wave solutions of \eqref{eq-sgKdV} and \eqref{eq-sgKdVu} for certain prescribed density distributions to illustrate the change of the corresponding wave profiles with depth.
 Moreover, a non-constant density distribution $\rho$ entails a non-vanishing vorticity\footnote{The voricity of the dimensionless flow field is given by $\omega = u_z- \e w_x$.} of the flow field at leading order, cf.~\eqref{eq-eqsfarfield_e^0} and \eqref{eq-u-eta}. This is to be expected in view of the rotational nature of stratified flows, see \cite{Con11, Walsh09}. 
\end{remark}

\section{Explicit traveling wave solutions}
\label{s-tws}

In this section we present explicit traveling wave solutions of equations \eqref{eq-sgKdV} and \eqref{eq-sgKdVu}. We study the effects of the non-constant density on the $z$-structure of solutions in general and for two concrete examples of density distributions.
Traveling wave solutions of \eqref{eq-sgKdV} of the form $\eta(\xi,\tau) =\vp(\xi-c\tau) $ with wave speed $c>0$  satisfy the ordinary differential equation 
\begin{equation}
\label{eq-tweq}
-2(c+\o_0)\vp' +\nu(\r) \vp\vp' + \frac{1}{3} \vp'''=0,
\end{equation}
with $\nu(\r)$ defined in \eqref{eq-nu}. 

\subsection{Solitary traveling waves}
\label{s-solitary}
 Assuming decay at infinity we may integrate this equation twice using $\vp'$ as an integrating factor in the second integration. 
The solitary traveling wave equation then reads 
\begin{equation}
    (\vp')^2=(6(c+\o_0)-\nu(\r)\vp)\vp^2.
\end{equation}
We see that real solutions can exist only if $\vp \leq \frac{6(c+\o_0)}{\nu(\r)}$. Following \cite{DrazinJohnson} we integrate the equation by writing
\begin{equation*}
\int \frac{\mathrm d \vp}{\vp \sqrt{6(c+\o_0)-\nu(\r)\vp}} = \pm \int \,\mathrm d s,
\end{equation*}
where $s=\xi-c\tau$, and using the substitution $\vp =\frac{6(c+\o_0)}{\nu(\r)}\sech^2(\theta)$. We obtain that 
 the solitary wave solutions of \eqref{eq-sgKdV}  are given by
\begin{equation}
\label{eq-sech-sol}
    \vp(\xi-c\tau) = \frac{6(c+\o_0)}{\nu(\r)} \sech^2\left(\frac{\sqrt{6(c+\o_0)}}{2}(\xi-c\tau)\right).
\end{equation}
Notice that for constant $\rho =1$ and $\Omega_0 =0$ this reduces to the familiar $\sech^2$-solutions of the standard KdV~\eqref{eq-stdKdV}. 
\begin{remark}
\label{rem-signofnu}
We observe that $\r$ appears only as a multiplicative factor in front of the solution and hence any changes in $\rho$ affect only the amplitude, cf.~Fig.~\ref{fig_solwave_sgKdV_tallwide}. This is different from the dependence on the Coriolis parameter and the wave speed: as $c$ or $\o_0$ increase, the profiles become taller as well as narrower, see  \cite{GeyQui2018}. 
Moreover, observe that we have made no assumption on the sign of $\nu(\rho)$. Therefore, in principle our approach allows us to obtain solitary waves of elevation when this number is positive and waves of depression when it is negative, see the remark at the end of Example 2. 
\end{remark}

For the horizontal velocity component $u$ traveling wave solutions of the equation \eqref{eq-sgKdVu} of the form $u(\xi, z,\tau)=\phi(\xi-c\tau, z)$ can be obtained directly from the explicit traveling wave solutions of  \eqref{eq-sgKdV}  in view of the relation between the surface  and the velocity component given in  \eqref{eq-u-eta}. This relation provides an explicit measure for the attenuation of the amplitude with depth which is inversely proportional to  $\r(z)$, see also Fig.~\ref{fig_solwave_sgKdV_depth}.  Therefore, we obtain a traveling wave solution of \eqref{eq-sgKdVu} for each depth $z\in [0,1]$  given by
\begin{equation}
\label{eq-sech-sol-u}
    \phi(\xi-c\tau, z) =\r(z)^{-1} \,\vp(\xi-c\tau),
\end{equation}
where  $\vp$ is a solution of \eqref{eq-tweq}. 
We observe that the amplitude of the wave does not vanish at the bottom. 

\subsubsection*{Example 1}
\begin{figure}[h!]
    \centering
    \begin{subfigure}[b]{0.45\textwidth}
         \includegraphics[width=\textwidth]{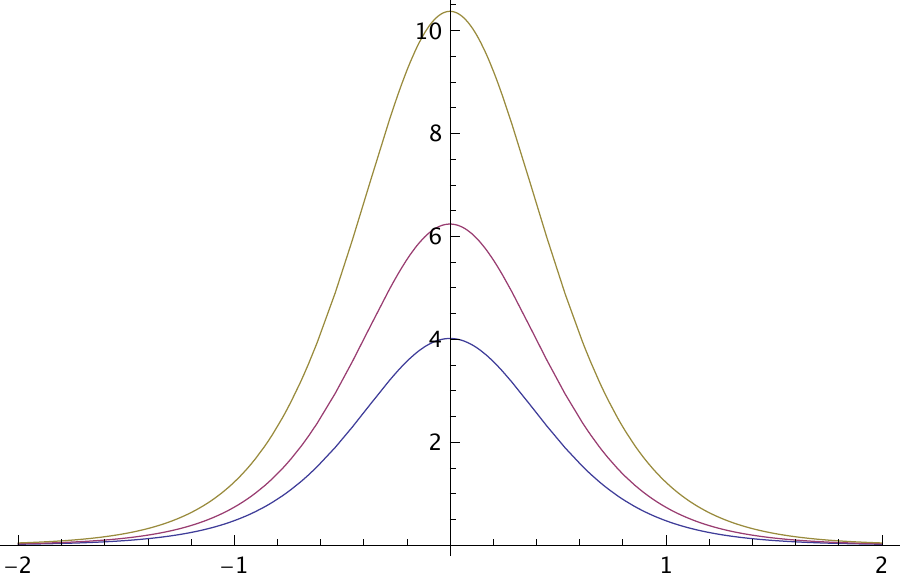}    
         \caption{}
         \label{fig_solwave_sgKdV_tallwide}
    \end{subfigure}
\begin{subfigure}[b]{0.5\textwidth}
\includegraphics[width=\textwidth]{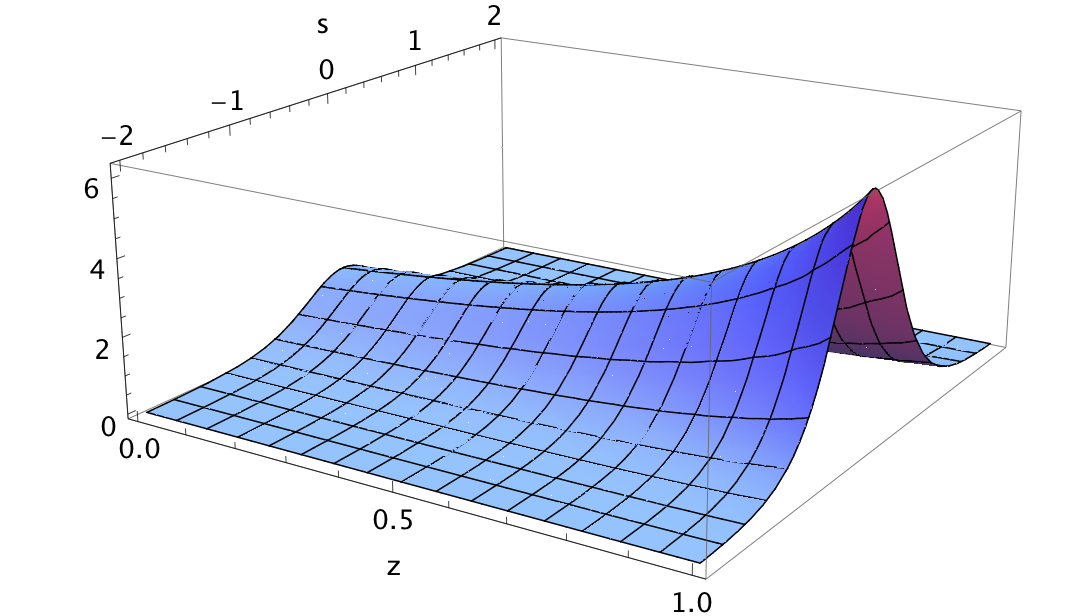} 
         \caption{}
         \label{fig_solwave_sgKdV_depth}
    \end{subfigure}
\caption{
Solitary traveling wave solutions \eqref{eq-sech-sol} of the surface equation \eqref{eq-sgKdV} with linear density function $\r(z)=1+A(1-z)$. In Fig.~\ref{fig_solwave_sgKdV_tallwide} we see that for larger  values of the parameter $A>0$ the profile becomes taller and also wider. In Fig.~\ref{fig_solwave_sgKdV_depth} we see how the amplitude of solutions decreases with depth $z\in [0,1]$ like $(Az)^{-1}$ according to \eqref{eq-sech-sol-u}.
}
\label{fig_solwaveKdV}
\end{figure}

Consider the linear density function 
\begin{equation}
\label{eq-linrho}
    \r(z) = 1 + A(1-z), \quad z\in[0,1]
\end{equation}
where $A>0$, i.e.~$\r(z)$ is increasing with depth. In this case the coefficient of the nonlinear term in the equation \eqref{eq-sgKdV} is $\nu(\r)=2\ln(1+A)/A +1 \in (1,3)$, which is a strictly monotonically decreasing function of $A$. Therefore,  for every fixed $z\in[0,1]$, the profiles of the solitary traveling wave solutions become taller and also wider as $A$ increases, cf.\eqref{eq-sech-sol} and  Fig.~\ref{fig_solwave_sgKdV_tallwide}.  The amplitude of the velocity component $\phi(\xi-c\tau, z)$ decreases with depth like $(Az)^{-1}$ in view of \eqref{eq-u-eta} and \eqref{eq-linrho}, see Fig.~\ref{fig_solwave_sgKdV_depth}. 

\subsubsection*{Example 2}
\begin{figure}[h!]
    \centering
    \begin{subfigure}[b]{0.45\textwidth}
    \includegraphics[width=.95\textwidth]{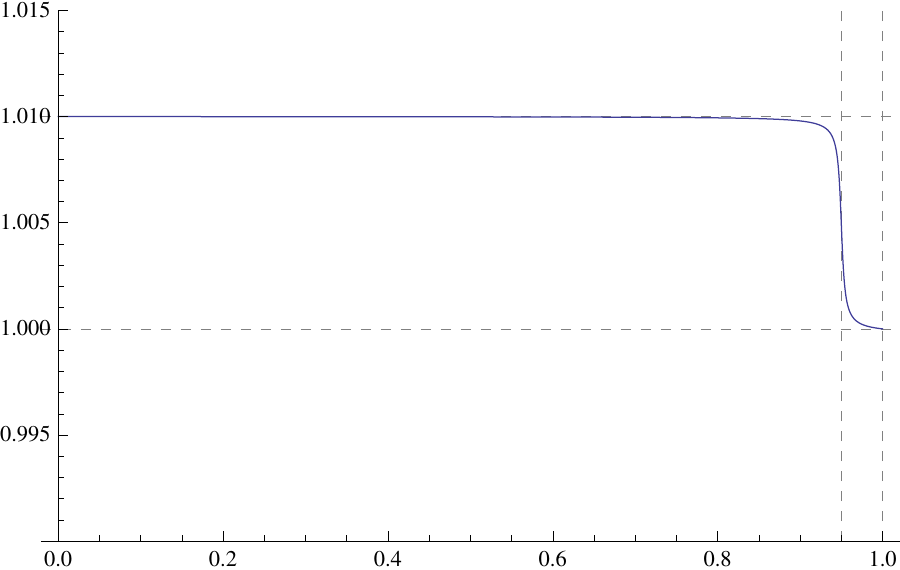}
        \caption{}
         \label{fig_rho_arctan}
    \end{subfigure}
\begin{subfigure}[b]{0.5\textwidth}
    \includegraphics[width=.95\textwidth]{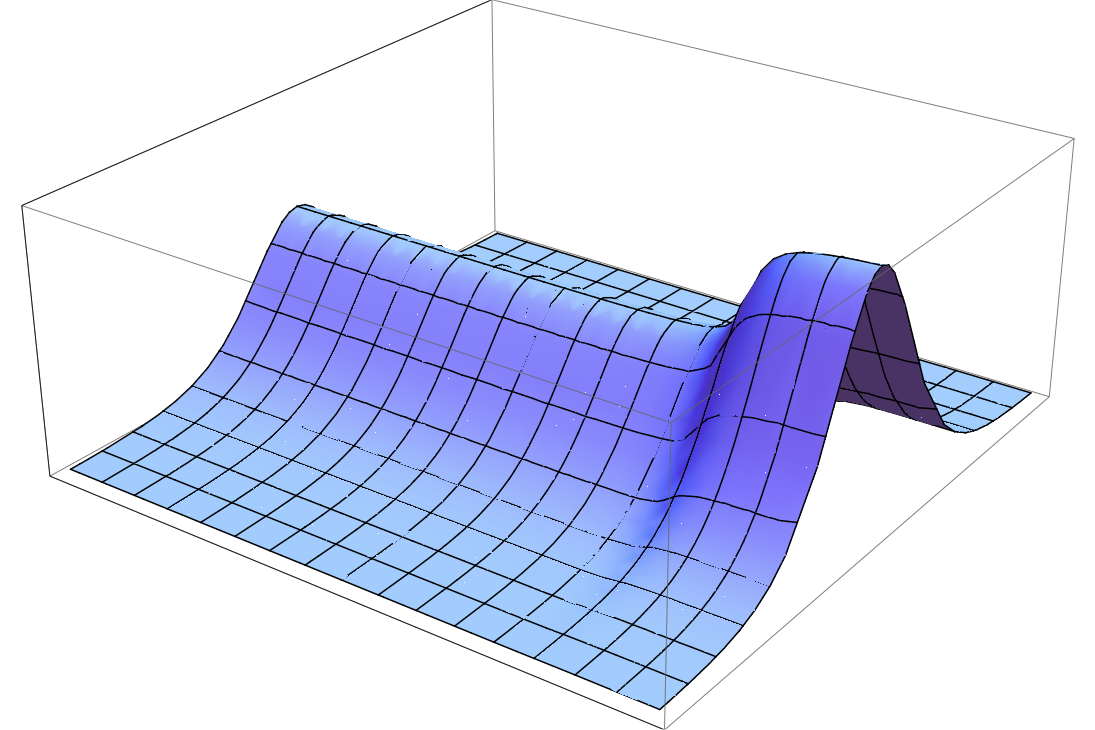}
         \caption{}
         \label{fig_sech2_arctan_depth}
    \end{subfigure}
\caption{
In Fig.~\ref{fig_rho_arctan} we see a plot of the density profile  $\r(z) = a_0-a_1\arctan\left(a_2(z-a_3)\right)$ defined in \eqref{eq-arctanrho}  for suitable choices of parameter values $a_i\in\R$ to model a density increase of $1\%$ from surface to bed.
Fig.~\ref{fig_sech2_arctan_depth}  shows a schematic representation of the fact that the amplitude of solitary wave solutions $\phi(\xi-c\tau, z)$  decays with depth inversely proportional to $\rho(z)$, cf.~\eqref{eq-sech-sol-u}.}
\label{fig_arctan_depth}
\end{figure}

Next we consider the density function
\begin{equation}
\label{eq-arctanrho}
    \r(z) = a_0-a_1\arctan\left(a_2(z-a_3)\right), \quad z\in[0,1],
\end{equation}
for suitable parameters $a_i\in\R$, to obtain a more realistic density distribution as described in the introduction and depicted in Fig.~\ref{fig:density}. We choose the parameters to match the actual physical situation in the mid-equatorial Pacific, with an average  water depth of about $4000\,m$  and a pycnocline located at a depth of roughly $200\,m$. In this oceanic region, the  density of the water at the surface is about $1027\,kg/m^3$ with a density  increase of $1\%$ from the surface to the seabed, cf.~\cite{CoJo15}. Nondimensionalizing and scaling these quantities to depths $z\in[0,1]$ and $\rho(1)=1$ leads us to the following choice of parameter values: $a_3=0.95$ determines where the region of biggest density gradient is located, i.e.~at the center of the pycnocline;  $a_2=300$ regulates the thickness of the pycnocline; with the last two parameters $a_0=1.0047$ and $a_1=0.0034$ we ensure the correct density increase. This choice results in the  density distribution depicted in Fig.~\ref{fig_rho_arctan}. 

It is interesting to see how this density profile is reflected in the decay of the amplitude of the solitary wave profile \eqref{eq-sech-sol} with depth. This is because the solitary waves $\phi(\xi-c\tau,z)$ of the velocity equation \eqref{eq-sgKdVu}  decay with depth inversely proportional to $\rho(z)$ according to relation \eqref{eq-sech-sol-u}, see Fig.~\ref{fig_sech2_arctan_depth}\footnote{Note that, since  the density change of $1\%$ in $\rho$ and the resulting impact on the amplitude decay in depth is quite small compared to the amplitude of the solitary wave, we have used a larger value for $a_1$ to produce Fig.~\ref{fig_sech2_arctan_depth} to make this effect more visible.}. 

Finally, we observe that for the choice of parameters described above  the coefficient of the nonlinear term  $\nu(\rho) = 2.97$ defined in \eqref{eq-nu} is positive. We note that in this setting negative values of $\nu(\rho)$, which give rise to solitary waves of depression, would arise only for a density increase of almost $300\%$ which is not a physically relevant scenario.

\subsection{Periodic traveling waves}
\label{s-periodic}
Periodic traveling waves of \eqref{eq-sgKdV} can be obtained in a similar fashion, cf.~\cite{DrazinJohnson} or \cite{Dingemans1997}. 
To this end we integrate \eqref{eq-tweq} twice to obtain 
\begin{equation}
\label{eq-twper}
    (\vp')^2=F(\vp) \quad \text{ where } \quad F(\vp) \coloneqq -\nu(\r)\vp^3 + 6(c+\o_0)\vp^2 +A \vp + B,
\end{equation}
where $A, B$ are constants of integration. One can show that this equation has real and bounded solutions only if $F(\vp)$ has three simple roots satisfying $\vp_3<\vp_2<0<\vp_1$ when $\nu(\rho)>0$ (the other situation being similar), in which  case we can write 
\begin{equation*}
F(\vp) = -\nu(\rho) (\vp-\vp_1) (\vp-\vp_2) (\vp-\vp_3).
\end{equation*}
The solution oscillates between  $\vp_2\leq \vp \leq \vp_1$ with period $2\int_{\vp_2}^{\vp_1}\frac{d\vp}{\sqrt{F(\vp)}}$. Therefore we can express the solution implicitly as 
\begin{equation*}
    s= s_1 \pm \int_{\vp}^{\vp_1}\frac{dg}{\sqrt{F(g)}},
\end{equation*}
where $\vp(s_1)=\vp_1$.  Using the substitution $g=\vp_1 - (\vp_1-\vp_2)\sin^2(\theta)$ this can be transformed into the standard elliptic integral
\begin{equation*}
    (s-s_1)\frac{\sqrt{\nu(\rho)(\vp_1-\vp_3)}}{2} = \pm 
    \int_{0}^{\phi}\frac{d\theta}{\sqrt{1-m\sin^2(\theta)}} \eqqcolon \pm v,
\end{equation*}
with elliptic modulus $m=\frac{\vp_1-\vp_2}{\vp_1-\vp_3}$ such that $m\in(0,1)$  and $\vp = \vp_2 + (\vp_1-\vp_2)\cos^2(\phi)$. Finally, using Jacobi elliptic functions and the fact that ${ \rm cn}(v|m)=\cos(\phi)$ is an even function, we obtain the  periodic cnoidal solution 
\begin{equation}
\label{eq-cnoidalsol}
        \vp(s) = \vp_2 + (\vp_1-\vp_2){\rm cn}^2
        \left(  (s-s_0)\frac{\sqrt{\nu(\rho)(\vp_1-\vp_3)}}{2}\bigg|\frac{\vp_1-\vp_2}{\vp_1-\vp_3}\right)
\end{equation}
in terms of the three distinct roots of $F(\vp)$. These roots are related to the parameters of the traveling wave equation \eqref{eq-twper}  via  the nonlinear relations
\begin{align*}
    \vp_1+\vp_2+\vp_3 = \frac{6(c+\o_0)}{\nu(\rho)}, 
    \quad \vp_1\vp_2+\vp_1\vp_3+\vp_2\vp_3 =\frac{-A}{\nu(\rho)}, 
    \quad \vp_1\vp_2\vp_3 =\frac{B}{\nu(\rho)},
\end{align*}
which one can solve as needed for concrete choices of density distributions $\rho$. We see from the explicit expression \eqref{eq-cnoidalsol} for periodic cnoidal wave solutions  that both the amplitude   $\frac{1}{2}(\vp_1-\vp_2)$   and the wave length $4K(m)(\nu(\rho)(\vp_1-\vp_3))^{-1/2}$, where $K(m) = \int_{0}^{\pi/2}\frac{d\theta}{\sqrt{1-m\sin^2(\theta)}}$, depend on the density distribution $\rho$ in an intricate way.  As in the case of solitary waves, explicit periodic traveling wave solutions of the velocity equation \eqref{eq-sgKdVu} can be obtained directly from \eqref{eq-cnoidalsol} in view of relation \eqref{eq-u-eta} for each depth $z\in[0,1]$.  We refrain here from any further qualitative discussion on periodic waves, which can be carried out along the lines of our analysis for solitary waves as illustrated in Section \ref{s-solitary}.

\subsection{Stability}
\label{s-stability}
Having established existence and some qualitative properties of traveling waves, we may ask about the stability of these solutions. The orbital stability of periodic and solitary traveling waves of the standard KdV have been studied by many authors, see for instance \cite{Benjamin1972b, Bona1987b, Deconinck2010}. Recall the observation presented in Remark \ref{rem-transftostdKdV} that for every fixed depth $z\in[0,1]$ the KdV type equations \eqref{eq-sgKdV} and \eqref{eq-sgKdVu}  derived  in Section \ref{S-KdV} can be transformed into the standard KdV. For this reason, we infer that the explicit solitary and periodic traveling wave solutions \eqref{eq-sech-sol} and \eqref{eq-sech-sol-u} of  equations \eqref{eq-sgKdV} and \eqref{eq-sgKdVu}, respectively,  are orbitally stable as well.

\subsection*{Acknowledgments}
Both authors acknowledge the support of the Erwin Schr\"odinger International Institute for Mathematics and Physics (ESI) during the program ``Mathematical Aspects of Physical Oceanography''. R.~Quirchmayr acknowledges the support of the European Research Council, Consolidator Grant No.~682537. Finally, we thank the reviewers for useful suggestions. 


\bibliographystyle{abbrv}

\end{document}